\documentclass[twocolumn, 10pt]{article}

\usepackage{geometry}
\usepackage{amssymb}
\usepackage[utf8]{inputenc}
\usepackage{hyperref}
\PassOptionsToPackage{hyphens}{url} % Allow URL hyphenation
\usepackage{url}
\usepackage{color}
\usepackage{xcolor}
\usepackage{pifont}
\usepackage{graphicx}
\usepackage{listings}
\usepackage[most]{tcolorbox}
\usepackage{tikz}
\usepackage{fancyvrb}
\usetikzlibrary{positioning,shapes, arrows,decorations.pathmorphing,mindmap,trees}

\definecolor{pcbgreen}{HTML}{0d9439}
\definecolor{redforti}{HTML}{e1251b}

\newtcbox{\mylink}[1][green]{on line,
  arc=0pt,outer arc=0pt,colback=#1!10!white,colframe=#1!50!black,
  boxsep=0pt,left=1pt,right=1pt,top=2pt,bottom=2pt,
  boxrule=0pt,bottomrule=1pt,toprule=1pt,
  breakable,
  left skip=0pt, right skip=0pt, width=0.98\linewidth % Allows better wrapping
}

\newcommand{\styledurl}[1]{%
    \mylink{\url{#1}}
}
\newcommand{\xmark}{\textcolor{red}{\ding{56}}}
\newcommand{\goldstar}{\textcolor{yellow!50!orange}{\ding{72}}}

\hypersetup{
    colorlinks=true,    % Enable colored links
    urlcolor=black      % Set link color to black (or your choice)
}
\title{Malware analysis assisted by AI with R2AI}
\author{Axelle Apvrille\thanks{Fortinet. \texttt{aapvrille@fortinet.com}} \and Daniel Nakov\thanks{daniel.nakov@gmail.com}}

\begin{document}
\maketitle

\section{Abstract}

This research studies the quality, speed and cost of malware analysis assisted by artificial intelligence. It focuses on Linux and IoT malware of 2024-2025, and uses r2ai, the AI extension of Radare2's disassembler. Not all malware and not all LLMs are equivalent but the study shows excellent results with Claude 3.5 and 3.7 Sonnet. 

\begin{itemize}
\item Despite a few errors, the quality of analysis is overall equal or better than without AI assistance. For good results, the AI cannot operate alone and must constantly be guided by an experienced analyst.
\item The gain of speed is largely visible with AI assistance, even when taking account the time to understand AI's hallucinations, exaggerations and omissions.
\item The cost is usually noticeably lower than the salary of a malware analyst, but attention and guidance is needed to keep it under control in cases where the AI would naturally loop without showing progress.
\end{itemize}

\section{Introduction}

Initially, malware analysis was performed exclusively through meticulous manual review of its assembly instructions. This approach, known as \textit{static analysis}, is very accurate but it is also unfortunately time-consuming. As the volume of malware increased, new techniques had to be developed to improve efficiency. Disassemblers such as IDA Pro (1996) and Radare2 (2006) improved. Dynamic analysis techniques were introduced, such as sandboxing and API hooking. It allows analysts to quickly extract key information (e.g  which remote server a malware sample contacts), but some of its drawbacks are precision and code coverage.

In practice, static analysis is still widely used for complex malware or those introducing novel techniques. Despite new disassembly tools (Binary Ninja in 2016, Ghidra in 2019) or improvements in the others, fully understanding a malware sample can take anywhere from one to ten days, depending on its complexity and originality.

Generative artificial intelligence (GenAI) is particularly suited for solving complex problems and integrating large volumes of data. Consequently, this raises the following thought: \textbf{can artificial intelligence (AI) speed up malware analysis while maintaining a level of quality comparable to that of human experts?}

In this paper, we study the results of AI's assistance on Radare2, or more precisely its disassembler, r2. 

To assess the results of AI assistance, we compare the analysis reports of independent researchers on given IoT and Linux malware, with the analysis we get with AI assistance. Malware evolving rapidly, the samples we select for this study are all recent: mid 2024 to early 2025. While this research focuses specifically on IoT and Linux malware, the methodology and findings may be applicable to Windows or Mac OS malware analysis. However, when analyzing mobile platforms such as Android, iOS or cross-platform frameworks like Flutter, different approaches might be necessary, potentially yielding different results.

The paper is structured as followed. First, we discuss related art and how AI is used for malware analysis in other circumstances. Then, in section \ref{sec:r2ai}, we introduce r2 and its AI extension, r2ai. In section \ref{sec:communication}, we focus on how r2ai communicates with the language model providers. Finally, in section \ref{sec:costs} and \ref{sec:results}, we discuss costs and quality.

\section{Related art}

The most straight forward way to query assistance from AI is a direct question to the AI, either through an API or using an online web chat interface. For example, OpenAI offers a free web chat where users can ask their questions. Unfortunately, binary upload is usually restricted, whether for size or security reasons. So, the end-user needs to copy/paste assembly code along with the question, which is not practical because, by nature, the assembly language is very precise and expresses only few actions per line. Pasting large amount of lines, moreover with no standard formatting, makes the conversation difficult to follow. This explains the need for a tool that interacts with a disassembler and an AI.

Several extensions exist for major disassemblers: for example SideKick \cite{sidekick} for Binary Ninja, IDA-Assistant \cite{ida-assistant} for IDA Pro and GptHidra \cite{gpthidra} for Ghidra. Their most common downside, compared to r2ai, is that they completely hide the prompts which are sent to the AI, making the interaction with the AI difficult to customize. Other common downsides are being tied to a given LLM (e.g IDA-Assistant only supports Anthropic Claude models), limited interactions with the disassembler (e.g GptHidra acts as an AI sidebar but does not modify the decompiled code), no ability to automatically run scripts for example for string deobfuscation.

Finally, some other projects explore different uses. BinaryChat \cite{binarychat} is used to spot vulnerabilities in C source code, especially in a CTF context. Reveng.AI \cite{reveng} focuses on source code, from a developer perspective: find vulnerabilities in software, create Yara rules for threat hunting etc. GhidraMCP \cite{ghidra-mcp} creates AI usable \textit{tools} for Ghidra - a concept which is similar Radare2 tools as we'll see in Section \ref{sec:auto-mode}. But, in its current implementation, GhidraMCP tools are limited to fine grained tasks (e.g. renaming functions) and do not act globally on improving clarity of the decompiled code.

\begin{table*}[t]
  \begin{tabular}{|p{2cm}|p{1.5cm}|c|p{1.5cm}|p{1.5cm}|c|p{2cm}|p{2cm}|}
    \hline
    \textbf{Project} & \textbf{Models} & \textbf{Decompile} & \textbf{Explain code} & \textbf{Spot vulnerabilities} & \textbf{Run scripts} & \textbf{Generate disassembler plugins} & \textbf{Customize AI prompt} \\
    \hline
  Aidapal & Custom mistral-based & \checkmark &  &  & &  &  \\
  \hline
  BinaryChat & OpenAI &  &  & \checkmark &  &  & \\
  \hline
  
  IDA-Assistant & Claude & \checkmark & ? & ? &  & &  \\
  \hline
  GptHidra & OpenAI &  & \checkmark &  &  &  & \\
  \hline
  GhidrAssist & Local models & \checkmark with limits & \checkmark & ? &  &  & \\
  \hline
  GhidraMCP & Several & \checkmark & To implement & To implement & \checkmark & & To implement \\
  \hline
  Reveng.AI & Custom model &  & ? & \checkmark &  &  & \\
  \hline
  Sidekick & Several & \checkmark & ? &  & Limited & \checkmark &  \\
  \hline
  \textbf{R2ai} & Several & \checkmark & \checkmark & \checkmark & \checkmark & \checkmark & \checkmark\\
  \hline
\end{tabular}
\label{tab:related-art}
\caption{Comparison of AI-assisted reverse engineering tools}
\end{table*}

\section{r2ai}
\label{sec:r2ai}

Radare2 \styledurl{https://www.radare.org} is a set of command-line tools for reverse engineering and binary analysis. Its primary tool known as \textbf{r2} is an interactive repl that can easily be scripted with different languages, natively supporting Javascript, and it supports many different architectures and binary file formats, in addition it is widely used by researchers to inspect binaries.

Late 2023, the idea to assist r2 with Artificial Intelligence emerged. A repository, named \textbf{r2ai} \styledurl{https://github.com/radareorg/r2ai}, was created and groups experimental tools related to assisting r2 with AI.

In 2025, based on prior experience with each of these tools, the project is being re-implemented to take its final shape, where r2ai is implemented as a \textit{plugin of r2}. End-users are expected to install r2ai as r2 \textit{package}, using the \texttt{r2pm} utility, then launch r2 on their binary, use r2 as they wish for their analysis and invoke AI assistance through commands prefixed by \texttt{r2ai} (see Figure \ref{fig:r2ai-cmd}). The implementation is performed in C for native integration with r2 and resource efficiency.

\begin{figure}[h]
  \footnotesize
\begin{lstlisting}[breaklines=true]
[0x000061d0]> r2ai -h
Usage: r2ai   [-args] [...]
| r2ai -d                 Decompile current function
| r2ai -dr                Decompile current function (+ 1 level of recursivity)
| r2ai -a [query]         Resolve question using auto mode
| r2ai -e                 Same as '-e r2ai.'
| r2ai -h                 Show this help message
| r2ai -i [file] [query]  read file and ask the llm with the given query
| r2ai -m                 show selected model, list suggested ones, choose one
| r2ai -n                 suggest a better name for the current function
| r2ai -r                 enter the chat repl
| r2ai -L                 show chat logs (See -Lj for json)
| r2ai -L-[N]             delete the last (or N last messages from the chat history)
| r2ai -R                 reset the chat conversation context
| r2ai -Rq ([text])       refresh and query embeddings (see r2ai.data)
| r2ai -s                 function signature
| r2ai -x                 explain current function
| r2ai -v                 suggest better variables names and types
| r2ai -V[r]              find vulnerabilities in the decompiled code (-Vr uses -dr)
| r2ai [arg]              send a post request to talk to r2ai and print the output
\end{lstlisting}
\caption{Supported commands by r2ai plugin on April 8, 2025}
\label{fig:r2ai-cmd}
\end{figure}

R2ai is not tied to a specific LLM: many models are supported and virtually any model which uses a supported API can be used. Indeed, a model is referenced by its API (or provider), e.g. \textit{openai, mistral, anthropic...}, and its specific model name e.g. \textit{mistral-large-latest}. R2ai may use models that run on their own proprietary server (e.g. ChatGPT), or open source servers such as Ollama \styledurl{https://ollama.com}. With Ollama, models can typically run locally, which may be important for confidentiality reasons for example.

\begin{table}
\begin{tabular}{c|p{5cm}}
  API & Model names \\
  \hline
  anthropic & claude-3-7-sonnet-20250219, claude-3-5-sonnet-20241022, claude-3-haiku-20240307 \\
  gemini & gemini-1.5-flash, gemini-1.0-pro \\
  groq & deepseek-r1-distill-llama-70b, deepseek-r1-distill-qwen-32b, llama-3.3-70b-versatile\\
  mistral & mistral-large-latest \\
  ollama & any LLM installed on Ollama \\
  openai & gpt-4, gpt-4o-mini, gpt-3.5-turbo \\
  xai & grok-2-1212\\
\end{tabular}
\caption{Non exhaustive list of supported LLMs}
\end{table}

There are r2ai commands:

\begin{enumerate}

\item To setup the model and its configuration. All configuration settings are accessible via \texttt{-e}. For example, \texttt{r2ai -e api=anthropic} and \texttt{r2ai -e model=claude-3-7-sonnet-20250219} (a shortcut exists for the latter: \texttt{r2ai -m claude-3-7-sonnet-20250219}). If the model requires an API key, this key is supplied to r2 via an environment variable. There are many configuration settings such as the maximum tokens to send, the ``temperature'' of the model (this controls its creativity), the system prompt, the expected output programming language for source code etc.
\item To query the AI. The end-user can ask his/her own questions on the code e.g. ``\textit{which URL does this binary contact?}''. Some common queries have their shortcut: \texttt{-d} to decompile (with AI assistance) a function, \texttt{-n} to suggest an appropriate name to the function, \texttt{-x} to explain the current function...
\end{enumerate}

\begin{figure}[h]
\begin{tikzpicture}
  \node (nR2AI) [rectangle, fill=blue!50!black, text=white, minimum width=2cm] at (0,0) {r2ai};
  \node (nLLM) [rectangle, fill=blue, text=white, right=2.5cm of nR2AI] {AI};
  \node (nR2) [rectangle, fill=black, text=white, left=2.5cm of nR2AI] {r2};
  \node (nAnalyst) [below=1.5cm of nR2AI] {\includegraphics[width=0.8cm]{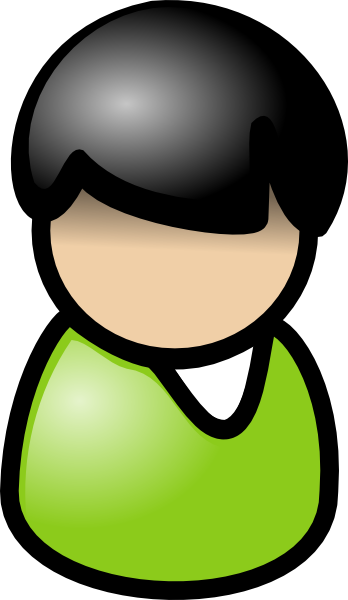}};
  \draw [->, ultra thick, color=blue] (1,0) -- (3.5,0);
  \draw [->, ultra thick, color=blue] (3.5,-0.5) -- (1,-0.5);
  \draw [->, ultra thick, color=blue] (1,-1) -- (3.5,-1);
  \draw [->, ultra thick, out=270, in=0, color=red!80!black] (nR2AI) to (nAnalyst);
  \draw [->, ultra thick, out=180, in=270, color=red!80!black] (nAnalyst) to (nR2AI);
  \draw [->, ultra thick] (-1.2,0) -- (-3.5,0);
  \draw [->, ultra thick] (-3.5,-0.5) -- (-1.2,-0.5);
  \node (n1) at (-2.5, -1) {r2 commands};
  \node (n2) [color=red!80!black] at (2, -3) {approve commands};
  \node (n3) [color=blue] at (2.5, -1.5) {system and user prompts};
\end{tikzpicture}
\caption{Requests and responses in Auto mode. It begins with an initial request from r2ai to AI.}
\label{fig:auto-mode}
\end{figure}

R2ai acts as a bridge between r2 and the AI: see Figure \ref{fig:auto-mode}, r2ai sits between r2 and AI\footnote{At implementation level, r2ai is a plugin of r2.}.

\section{r2ai communications}
\label{sec:communication}
\subsection{Direct mode}

There are two different ways to communicate with the AI: the \textit{direct} mode and the \textit{auto} mode. In the direct mode, the end-user's prompt is directly forwarded to the AI. The prompt and additional data depend on the requested action. For example, function decompilation (\texttt{r2ai -d}) adds the function pseudo code (close to assembly) to the context (see Figure \ref{fig:r2ai-decompile}).

\begin{figure}[h]
\begin{lstlisting}[breaklines=true]
curl -X POST \
https://api.anthropic.com/v1/messages \
-H 'anthropic-version: *****' -H 'x-api-key: **************' -H 'accept: *****' -H 'content-type: *****' \
-d '{'model': 'claude-3-7-sonnet-20250219', 'messages': [{'role': 'user', 'content': [{'type': 'text', 'text': 'Explain prctl in 1 line'}]}], 'temperature': 0.002, 'top_p': 0.95, 'max_tokens': 4096}'
\end{lstlisting}
\caption{Example of r2ai direct request}
\label{fig:r2ai-direct}
\end{figure}

 \begin{figure}[h]
 \begin{lstlisting}[breaklines=true]
curl -s https://api.mistral.ai/v1/chat/completions -H "Authorization: Bearer XXXXXXXXXXXXXX" -H "Content-Type: application/json" -d '{"stream":false, "model":"codestral-latest", "messages" :[{"role":"user", "content":"Rewrite this function and respond ONLY with code, NO explanations, NO markdown, Change 'goto' into if/else/for/while, Simplify as much as possible, use better variable names, take function arguments and strings from comments like 'string:', Convert this pseudocode into C\nRewrite this function and respond ONLY with code, NO explanations, NO markdown, Change goto into if/else/for/while, Simplify as much as possible, use better variable names, take function arguments and strings from comments like string:. Translate this code into C programming language. Do not explain anything:\nOutput of pdc:\n[BEGIN]\n// callconv: x0 arm64 (x0, x1, x2, x3, x4, x5, x6, x7, stack);\nvoid entry0 (int64_t arg1, int64_t arg_0h, int64_t arg_8h)...
 \end{lstlisting}
 \caption{Example where the end-user issued command \texttt{r2ai -d} which decompiles a given function. In this particular case, the LLM was Mistral's codestral-latest. R2ai provides to the AI the function's pseudo code (output of r2 command \texttt{pdc}). An API key is used to access Mistral via the authorization header. Based on this context, the AI is expected to answer with corresponding code in C.}
 \label{fig:r2ai-decompile}
 \end{figure}

The AI deals with what it receives and answers back with the solution. If the end-user asks another question, the question piles up in the context (Figure \ref{fig:direct-context}). To save tokens and reduce costs, it is important to \textit{reset} the context when a new conversation begins (option \texttt{-R}).

\begin{figure}[h]
  \centering
\begin{tikzpicture}
  \node (n1) [rectangle, fill=gray!20, minimum width=6cm, text width=6cm] at (0,0) {'role': 'user', 'content': 'question 1'};
  \node (n2) [rectangle, fill=gray!20, minimum width=6cm, text width=6cm, below=0.2cm of n1] {'role': 'user', 'content': 'question 2'};
  \node (n3) [rectangle, fill=gray!20, minimum width=6cm, text width=6cm, below=0.2cm of n2] {'role': 'user', 'content': 'question 3'}; 
\end{tikzpicture}
\caption{Questions pile in the context sent to the AI}
\label{fig:direct-context}
\end{figure}

\subsection{Auto mode}
\label{sec:auto-mode}

R2ai features an ``automatic'' mode via the command \texttt{r2ai -a}. 
In the \textit{auto} mode, the AI may use 4 different tools provided by r2ai:

\begin{enumerate}
\item \textbf{r2cmd}. A tool for the AI ask execution of a r2 command, such as \texttt{pdf main} which decompiles the main function.
\item \textbf{execute\_binary}. The AI asks execution of a given binary. This binary is expected to be on the end-user's host and will be executed on the end-user's host.
\item \textbf{run\_python}. Runs a Python program on the end-user's host. Creates a process by launching \texttt{python -c generatedprogram.py}. Python is expected to be installed on the end-user's host.
\item \textbf{execute\_js}. Runs a Javascript program, using \texttt{QuickJS} engine which comes built into Radare2 distributions.
\end{enumerate}

In a first request to the AI, r2ai sends:

\begin{enumerate}
\item a \textbf{System prompt}. This prompt explains the context to the AI: \textit{``You are a reverse engineer and you are using radare2 to analyze a binary. The user will ask questions...}.
\item \textbf{Initial information}. Typically, this contains information on the binary and the list of all functions r2 found. This is customizable in r2ai by an option named \textit{auto.init\_commands}, which sets the initial r2 commands to launch and include in the first request.
\item the \textbf{User prompt}. This is the question the AI needs to address.
\item a \textbf{Definition of available tools} (example at Figure \ref{fig:tools}). It explains what the AI may use to answer the question at best.
  \end{enumerate}
  
\begin{figure}[h]
\begin{lstlisting}[breaklines=true]
'tools': [
  {'name': 'r2cmd',
   'input_schema': {'type': 'object', 'properties': {'command': {'type': 'string'}}, 'required': ['command']},
   'description': 'Run a r2 command and return the output'}
]
\end{lstlisting}
\caption{Definition of the r2cmd tool, sent in a context to Anthropic Claude 3.7 Sonnet}
\label{fig:tools}
\end{figure}

After the first request, the AI responds, possibly requesting use of a tool. In that case, r2ai asks the end-user to review the command, and if approved, the command is executed. If it's a r2cmd, the command is sent to r2 via r2pipe mechanism. The execution answer is added to the context that r2ai sends to the AI.

There may be multiple such interactions. At some point, the AI should hopefully have enough information to conclude, however by security in case the loop is endless, a limit is defined in the configuration key \textit{r2ai.auto.max\_runs}.

At the end, the AI is expected to finally answer with a response that no longer uses any tool and provides the answer to the end-user's initial prompt.

\begin{tcolorbox}[enhanced, attach boxed title to top center={yshift=-2mm}, colback=green!5!white, colframe=green!40!black,colbacktitle=green!40!black,title=How useful is the run\_python tool?]
  The tool is particularly useful to solve string obfuscation or decryption.
  \begin{itemize}
  \item The AI analyzes the binary and works out an algorithm to decrypt encrypted strings. It generates a Python program to decrypt them.
  \item The AI asks the Python program to be run using run\_python tool. R2ai asks the end-user for approval.
  \item If approved, the program runs on the end-user's hosts. The output of the program is sent back to the AI.
  \item In case of compilation or runtime errors, the AI is able to fix its program using the output, until it runs correctly and decrypts the strings.
  \end{itemize}
  
\end{tcolorbox}

Note the binary is never sent to the AI\footnote{In theory, binary data may be sent to the AI but that will only happen if the AI generates a program that does precisely that, then asks its execution through execute\_binary, and the end-user approves the execution of this binary.}. The AI only ever receives text information containing typically assembly instructions (text format), list of functions (text), list of strings (text) etc.

\subsection{MCP and user approval}

MCP (Model Context Protocol) \styledurl{https://modelcontextprotocol.io} is an open protocol that standardizes how applications provide context to LLMs. It is particularly useful to expose \textit{tools} the AI can use in a standard way. The concept is not new: r2ai started exposing tools to the AI before MCP existed. MCP standardized it.

The \textbf{r2mcp} \cite{r2mcp} project implements a Radare2 MCP server. For example, in ChatGPT desktop, end-users can launch a tool that lists imports or classes of a binary. This runs r2 with the appropriate command.

In r2ai, all tools run \textbf{on the end-user's host}: run\_python executes a Python program on the host etc. This is dangerous, even if the LLM is not considered infected or malicious: what if the AI asks to erase the disk, believing it is in a sandboxed environment? Consequently, r2ai systematically prompts the end-user for approval and modification of each command. This is fine-grained approval: the end-user can modify each r2 command and can edit any line of programs to execute.

MCP tools tend to query user approval in a global manner. For example, in \cite{ghidra-mcp} at 3 min 38 sec, when Claude AI queries a ghidra tool, a dialog pops up requesting approval to run ``search\_functions\_by\_name'' from Ghidra. The end-user can allow or deny, but cannot review the details nor edit them. The dialog will rule out obvious malicious MCP servers, but it leaves the door open to smarter abuses (a malicious MCP server convincing the end-user it is legitimate) and to use risks of the tool. A malware could intentionally exploit the MCP server tool, or as we said previously, a benign LLM can unfortunately run dangerous commands. \textbf{While fine grained approval, as implemented by r2ai, may seem cumbersome, it is the only way to go for malware analysis}.

\begin{tcolorbox}[enhanced, attach boxed title to top center={yshift=-2mm}, colback=red!5!white, colframe=red!80!black,colbacktitle=red!80!black,title=Benign LLMs may have risky behavior]
  During the analysis of a Linux malware, the AI requested the r2cmd tool to open a debug session (r2 commands \texttt{do, db, dc}...). This means \textit{executing} the malware on the analyst's host, and may result in host infection if malware runs until the infection routine. Fortunately, this was blocked by user review.
  This is a perfect example where the AI meant no harm, but its use of the exposed tools would have been dangerous.
  \end{tcolorbox}

\section{Costs}
\label{sec:costs}

\begin{figure}
  \includegraphics[width=8cm]{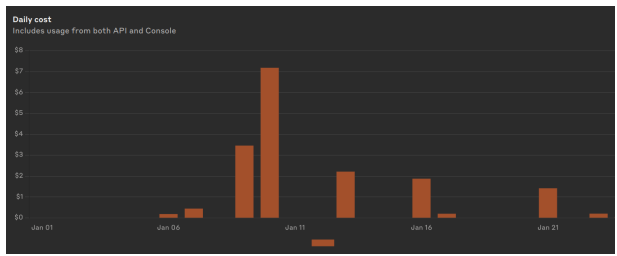}
  \caption{Cost difference between direct and automatic mode. January 6 and 7: only direct mode. January 9 and 10: automatic mode.}
  \label{fig:cost-money}
\end{figure}

Obviously, costs entirely depend on AI subscription.
Let's take the example of a malware analyst, whose job is to understand all main aspects of a given binary. We assume the analyst keeps focused all day long on this task, and s/he is either thinking, manually reversing or querying the AI. In such a case, queries through direct mode will cost at most 1 or 2 USD per day. The automatic mode, having usually a bigger context and several interactions, is more expensive and will easily reach 7 to 10 USD per day (or more if not controlled). See Figure \ref{fig:cost-money}.

The r2ai automatic mode shows costs by default in a status message (see Figure \ref{fig:cost-status}).

\begin{figure}[h]
\begin{lstlisting}
anthropic/claude-3-7-sonnet-20250219 | total:
  $0.0153540000 | run: $0.0153540000 | 1 / 100
  | 7s / 7s
\end{lstlisting}
\caption{Typical status message displayed during an automatic interaction with r2ai. In this case, there was only 1 interaction. The limit was set to 100 interactions. The cost was 0.015 USD}
\label{fig:cost-status}
\end{figure}

There are several ways to reduce costs:

\begin{enumerate}
\item Obvious solution: use a \textbf{free local model or free API key} e.g. Mistral currently offers free API keys. Alternatively, one might consider using a free model from a high performance cloud instance: for a team of analysts, the cost of the cloud instance (expect 2 euros per hour) may be lower than purchasing API keys for all (note that sharing the same API key among several users does not work well, because everybody hits token limits faster).

\item In r2ai auto mode, limit interactions (\texttt{-e r2ai.auto.max\_runs}). For binary analysis, in our experience, limiting to 15 interactions is more than enough. If the AI hasn't answered correctly by then, it's better to modify the prompt.
\item \textbf{Shorten the context} as much as possible. For example, by default, the output of \texttt{aaa;iI;afl} is included in each request in the automatic mode. Command \texttt{afl} lists all functions in the binary: the list can be very long for some binaries, it's a better idea to output only relevant functions (e.g \texttt{afl~main} searching just for the \texttt{main}).
\item \textbf{Reset the context} as soon as the conversation is finished (\texttt{-R}). Otherwise, all future questions pile up in the context, and each time, that's more tokens to send.
\item Use \textbf{token limits} that will truncate the context if necessary (\texttt{r2ai.max\_tokens}). Truncating the context may lead to issues if it's in the middle of something interesting, but in several cases, when the context is long, lots of information can actually be cut off.
\item Ask for concise answers in your \textbf{prompts}. For example, adding a mention like ``answer in 1 or 2 lines at most'' is efficient, and saves tokens, thus costs.

\end{enumerate}

\section{Results}
\label{sec:results}

\subsection{Quality of analysis}

For several IoT and Linux malware of 2024 and 2025 (see Table \ref{tab:ioc}), we compare the quality of analysis between various models (\ref{sec:quality-models}), and without AI assistance (\ref{sec:quality-human}).

\begin{table*}
  \begin{tabular}{l|l}
    IOC & Malware name \\
    \hline
    \texttt{43f72f4cdab8ed40b2f913be4a55b17e7fd8a7946a636adb4452f685c1ffea02} & Linux/Devura.A!tr \cite{devura}, aka Sedexp \\
    \texttt{712d9abe8fbdff71642a4d377ef920d66338d73388bfee542f657f2e916e219c} & ELF/Agent.JL!tr.dldr aka Goldoon \cite{goldoon} \\
    \texttt{89b60cedc3a4efb02ceaf629d6675ec9541addae4689489f3ab8ec7741ec8055} & Linux/RudeDevil.A!tr \cite{rudedevil-analysis} \\
    \texttt{e146baab13210b41abaec473c0b536b13b54fbf1a55489b093aa5215ac92c93b} & Linux/Ngioweb.A!tr \\
    \texttt{94e8540ea39893b6be910cfee0331766e4a199684b0360e367741facca74191f} & ELF/Sshdinjector.A!tr \cite{sshdinjector-ai} \\
    \texttt{943e1539d07eaffa4799661812c54bb67ea3f97c5609067688d70c87ab2f0ba4} & Linux/Ladvix.E!tr aka Rhombus, Ebola \\
    \texttt{fd8441f8716ef517fd4c3fd552ebcd2ffe2fc458bb867ed51e5aaee034792bde} & Linux/Shellcode\_ConnectBack.H!tr \\
    \texttt{cc7ab872ed9c25d4346b4c58c5ef8ea48c2d7b256f20fe2f0912572208df5c1a} & Linux/Prometei.B!tr \cite{prometei-ai} \\
  \end{tabular}
  \caption{List of malware we reversed using r2ai}
  \label{tab:ioc}
\end{table*}
    
\subsubsection{Quality of analysis per model}
\label{sec:quality-models}

To compare quality of code, we asked r2ai to decompile the main of the same sample of Linux/Shellcode\_ConnectBack.H!tr with various models:

\begin{itemize}
\item OpenAI ChatGPT 4.5 preview (Feb 2025)
\item Anthropic Claude 3.7 Sonnet 20250219
\item Codellama 70b (July 2024)
\item DeepSeek-r1 32b (Feb 2025)
\item Microsoft Phi4 14b (Jan 2025)
\item Alibaba Hhao qwen2.5-coder-tools 32b (Sept 2024)
\item IBM granite-code 34b (Sept 2024)
\item Mistral codestral-2502
\end{itemize}

Code quality is measured by general properties (e.g capability to rename appropriately variables) and by specific correctness for this sample (e.g. good recovery of the IP address the malware connects to). See Table \ref{tab:code-quality}.

\begin{table*}
\begin{tabular}{p{3.5cm}|c|c|c|c|c|c|c|c}
  Code quality & ChatGPT 4 & Claude 3.7 & Codellama & DeepSeek & Phi4 & Qwen 2.5 & Granite & Mistral \\
  \hline
  Renames variables and functions appropriately & \textcolor{pcbgreen}{\checkmark} & \textcolor{pcbgreen}{\checkmark} & \textcolor{pcbgreen}{\checkmark} & \textcolor{pcbgreen}{\checkmark} & \textcolor{redforti}{\xmark} & \textcolor{pcbgreen}{\checkmark} & \textcolor{redforti}{\xmark} & \textcolor{redforti}{\xmark} \\
  Comment quality & None & \goldstar \goldstar \goldstar & None & \goldstar \goldstar & None & None & \textcolor{redforti}{\xmark} & None \\
  Transforms constants to correct flag name &  \textcolor{pcbgreen}{\checkmark} & \textcolor{pcbgreen}{\checkmark} & \textcolor{pcbgreen}{\checkmark}  & \textcolor{redforti}{\xmark} & \textcolor{pcbgreen}{\checkmark} &\textcolor{redforti}{\xmark} & \textcolor{redforti}{\xmark} \\
  General readability & \goldstar \goldstar &  \goldstar \goldstar \goldstar & \goldstar \goldstar & \goldstar \goldstar \goldstar & \textcolor{redforti}{\xmark} & \goldstar \goldstar \goldstar & \textcolor{redforti}{\xmark} & \textcolor{redforti}{\xmark} \\
  Resolved system calls & \goldstar \goldstar & \goldstar \goldstar \goldstar & \goldstar \goldstar \goldstar & \goldstar \goldstar \goldstar & \goldstar & \goldstar \goldstar & \textcolor{redforti}{\xmark} & \textcolor{redforti}{\xmark} \\
  Correct IP address and port & \goldstar & \goldstar \goldstar & \textcolor{redforti}{\xmark} & \textcolor{redforti}{\xmark} & \goldstar & \textcolor{redforti}{\xmark} & \textcolor{redforti}{\xmark} & \textcolor{redforti}{\xmark} \\
  Correct mprotect call & \goldstar & \goldstar \goldstar & \textcolor{redforti}{\xmark} & \goldstar \goldstar \goldstar & \goldstar & \goldstar & \textcolor{redforti}{\xmark} & \textcolor{redforti}{\xmark} \\
  Code coverage (does not miss important points) & \goldstar \goldstar & \goldstar \goldstar \goldstar & \goldstar & \goldstar \goldstar \goldstar & \goldstar \goldstar \goldstar & \goldstar & \textcolor{redforti}{\xmark} & \goldstar \goldstar \\
  \hline
  Total & 10 & 18 & 6 & 13 & 5 & 6 & -8 & -3 \\
  \hline
\end{tabular}
\caption{Comparing the quality of result of \texttt{r2ai -d} on the entry point of Linux/Shellcode\_ConnectBack.H!tr. For the total rating, we count +1 for each star or checkmark and -1 for each red cross. Important: the intent of this table is to encourage users to try different LLMs on their samples if one is not satisfactory. The intent is \textit{not} to rate LLMs, as this would require tests on more samples.}
\label{tab:code-quality}
\end{table*}

From this test, we do not aim to accurately compare models: some models behave better on a given sample than on another, we'd need to test a representative set of samples and code quality for those by each model for such a conclusion. However, so far, in each sample we tried, we got our best results with Claude 3.5 or 3.7. Mistral, Qwen, ChatGPT and DeepSeek also gave interesting results in some case, useful to complement the output of Claude.

\subsubsection{Quality of analysis compared with a human only analysis}
\label{sec:quality-human}

For several IoT and Linux malware of 2024 and 2025 (see Table \ref{tab:ioc}), we compare the quality of analysis made by humans only, and with AI's assistance. The ``human-only'' analysis is done by independent researchers \cite{devura} \cite{goldoon} \cite{rudedevil-analysis}. The analysis assisted by AI was performed by us, using r2ai. 

We used various models, listed in Table \ref{tab:models-compare}, but mostly Anthropic Claude 3.5 and 3.7 Sonnet, and Mistral codestral 2502.

\begin{table*}
\begin{tabular}{l|l}
  Model name & Type \\
  \hline
  codegeex4:latest & Locally running on Ollama server \\
  deepseek-r1:latest & Locally running on Ollama server \\
  QuantFactory/granite-8b-code-instruct-4k-GGUF & Locally running on r2ai-server \\
  anthropic:claude-3-7-sonnet-20250219 & Remote access via paid API key \\
  anthropic:claude-3-5-sonnet-20241022 & Remote access via paid API key \\
  openai:gpt-4 & Remote access via paid API key \\
  openai:gpt-4o & Remote access via paid API key \\
  codestral-2502 & Remove access via free API key to Mistral \\
\end{tabular}
\caption{Models used to analyze Linux and IoT malware and compare quality of analysis}
\label{tab:models-compare}
\end{table*}

The detailed comparison for Linux/Devura, Goldoon and RudeDevil are provided at Tables \ref{tab:compare-devura}, \ref{tab:compare-goldoon} and \ref{tab:compare-rudedevil}. For Linux/Devura, the AI uncovered the format of arguments to the binary, explaining the malware could run Linux commands that way. This had not been mentioned by the ``human''-only analysis.

\begin{table}[h]
 \begin{tabular}{p{3cm}|c|c}
   & Only Human & AI Assistance \\
     \hline
     Persistance through udev rules & \textcolor{pcbgreen}{\checkmark} & \textcolor{pcbgreen}{\checkmark} \\
     Reverse shell & \textcolor{pcbgreen}{\checkmark} & \textcolor{pcbgreen}{\checkmark} \\
     Manipulate arguments & \textcolor{pcbgreen}{\checkmark} & \textcolor{pcbgreen}{\checkmark} \\
     Change process name & \textcolor{pcbgreen}{\checkmark} & \textcolor{pcbgreen}{\checkmark} \\
     Explain prctl and kdevtmpfs & & \textcolor{pcbgreen}{\checkmark} \\
     Remote IP address as argument & & \textcolor{pcbgreen}{\checkmark} \\
     \texttt{luv} command & & \textcolor{pcbgreen}{\checkmark} \\
     \hline
     \textbf{General} & \goldstar \goldstar \goldstar  & \goldstar \goldstar \goldstar \goldstar  \\
 \end{tabular}
 \caption{Comparing human analysis of Linux/Devura \cite{devura} with an analysis using r2ai. Checkmarks only account for success/failure of given tasks. They do not detail quality. For example, changing process names is reversed in higher quality with AI assistance than without. On the ``general'' line, the number of stars represents a global impression, not a count of check marks. }
 \label{tab:compare-devura}
\end{table}

For the analysis of Goldoon, the AI understood better the ``XOR'' algorithm mentioned by the human analysis. It's actually a little more than a simple XOR algorithm. The AI's code uncovers that. But, on the other side, the AI had a very hard time finding the correct URL the malware was talking to. The AI simplified the URL to \texttt{/bins/aarch64-linux-gnu}. This would have been \textit{logical}, the malware getting a URL, depending on the architecture name. In reality, the implementation is slightly different and the malware gets a URL \texttt{/bins/ENCRYPTED-ARCH}, where \texttt{ENCRYPTED-ARCH} is the encryption of the string \texttt{aarch64-linux-gnu}.

\begin{table}[h]
  \begin{tabular}{p{3.2cm}|c|c}
    & Only Human & AI Assistance \\
    \hline
    XOR algorithm & \goldstar & \goldstar \goldstar \\
    XOR key & \textcolor{pcbgreen}{\checkmark} & \textcolor{pcbgreen}{\checkmark} \\
    URL has encrypted value & \textcolor{pcbgreen}{\checkmark} & \textcolor{redforti}{\xmark} \\
    Uses fixed user agent & \textcolor{pcbgreen}{\checkmark} & \textcolor{pcbgreen}{\checkmark} \\
    Kill other instances & & \textcolor{pcbgreen}{\checkmark} \\
    Block specific signals &  & \textcolor{pcbgreen}{\checkmark} \\
    \hline
    \textbf{General} & \goldstar \goldstar \goldstar \goldstar & \goldstar \goldstar \goldstar \goldstar \\
  \end{tabular}
  \caption{Comparing human analysis of Goldoon \cite{goldoon} with an analysis using r2ai}
 \label{tab:compare-goldoon}
\end{table}

The analysis of Linux/RudeDevil was excellent with or without AI assistance. With AI assistance, we got more easily interesting details on which services and which configuration differences were made when root or not (file descriptor limits). Decrypting encrypted data was also achieved assisted by AI \cite{rudedevil-obfuscation} but it required several attempts and modifications of the Python decryptor the AI created. 

\begin{table}[h]
  \begin{tabular}{l|c|c}
    & Only Human & AI Assistance \\
    \hline
    Malware author message &  \textcolor{pcbgreen}{\checkmark} & \textcolor{pcbgreen}{\checkmark} \\
    Daemon &  \textcolor{pcbgreen}{\checkmark} & \textcolor{pcbgreen}{\checkmark} \\
    Signal handlers &  \textcolor{pcbgreen}{\checkmark} & \textcolor{pcbgreen}{\checkmark} \\
    Check for root &  \textcolor{pcbgreen}{\checkmark} & \textcolor{pcbgreen}{\checkmark} \\
    Starting services &  \textcolor{pcbgreen}{\checkmark} & \textcolor{pcbgreen}{\checkmark} \\
    Details of services & & \textcolor{pcbgreen}{\checkmark} \\
    Thread for mining &  \textcolor{pcbgreen}{\checkmark} & \textcolor{pcbgreen}{\checkmark} \\
    Decryption XOR based &  \textcolor{pcbgreen}{\checkmark} & \textcolor{pcbgreen}{\checkmark} \\
    File descriptor limits & & \textcolor{pcbgreen}{\checkmark} \\
    \hline
    \textbf{General} & \goldstar \goldstar \goldstar \goldstar \goldstar & \goldstar \goldstar \goldstar \goldstar \goldstar \\
  \end{tabular}
\caption{Comparing human analysis of Linux/RudeDevil \cite{rudedevil-analysis} with an analysis using r2ai}
 \label{tab:compare-rudedevil}
\end{table}

Globally, the results are the following:

\begin{itemize}
\item In terms of quality only, the results with AI assistance were similar or better than what was observed without AI. We will discuss analysis speed in \ref{sec:speed}. Quality of analysis depends on each sample: AI has been seen to perform better on some malicious samples (e.g Linux/Devura) than on others (Linux/Prometei) \cite{prometei-ai}.
\item Decompiled code, generated by AI, is easy to read and understand, with helpful comments and variable names.
\item Explanations from the AI were usually of excellent quality \cite{insomnihack}.
\item The AI is able to work out algorithms of easy to medium complexity and write correct code or decryptors. For example, the AI was able to solve string obfuscation of Linux/RudeDevil \cite{rudedevil-obfuscation}.
\item Unpacking - even well known unpacking such as UPX \styledurl{https://upx.github.io/} - is too difficult to solve for current LLMs \cite{prometei-ai}. Malware analysts must \textit{first unpack} the sample and \textit{then ask} for AI assistance on the unpacked version.
\item Several issues were noticed (hallucinations, exaggerations, omissions - see Section \ref{sec:typical-issues}). They are usually noticed by incoherent responses from the AI, but sometimes require experience and attention from the reverse engineer (e.g. side by side disassembly with and without AI).
\end{itemize}

  \subsection{Issues}
\label{sec:typical-issues}  

When using r2ai for reverse engineering, one may encounter several issues.

There are \textbf{benign issues} such as communication errors with the AI, rate limit errors or not enough credits.
Rate limits correspond to a maximum number of input tokens for a given period, e.g 40,000 input tokens per minute. Those issues are benign because they solve on their own with patience (or by purchasing the proper subscription plan).

There are \textbf{hard limits} like too big contexts. For example, at Figure \ref{fig:context-length}, the AI was sent 11,219 tokens, where the AI's limit is 8,192. This issue can be solved by changing the maximum input tokens limit... or purchasing a better subscription.

\begin{figure}[h]
\begin{lstlisting}
"message": "This model's maximum context
  length is 8192 tokens. However, you
  requested 11219 tokens (6091 in the
  messages, 5128 in the completion).
  Please reduce the length of the
  messages or completion.",
"type": "invalid_request_error",
"param": "messages",
"code": "context_length_exceeded"
\end{lstlisting}
\caption{Context length exceeded error.}
\label{fig:context-length}
\end{figure}

Finally, there are \textbf{analysis limits} by the AI. Depending on how adequately the AI was trained, it will offer different qualities of answers. The following unfortunately occur for any model so far:

\begin{itemize}
\item \textbf{Hallucinations} are the best known analysis issue (example at Figure \ref{fig:hallucination}). The AI invents something which is not true. AIs can be very convincing, so, sometimes, the hallucination can be difficult to spot. The only solution consists in checking what the AI says, either manually, or by asking again, if possible after resetting the context (\texttt{-R} in r2ai).
\item \textbf{Exaggerations} are another form of analysis issue. In this case, the AI does not totally invent, but it largely exaggerates the fact. For example, in Figure \ref{fig:manipulation}, the code prints a MAC address, and the AI reported that as ``MAC address manipulation code''.
  
\item Finally, there are \textbf{omissions}. Omissions are frequent, because AIs generally focus more on giving a correct overview of code, rather than providing every detail (Figure \ref{fig:omission}). As AI is not aware of our interests, it may forget to mention something a reverse engineer would find extremely interesting. There is no other solution than to ask again, with a more precise question.
\end{itemize}

\begin{figure}[h]
  \scriptsize
  \begin{verbatim}
case SERVER_REQ_FILE_DOWNLOAD:
  std::string file_path = getPacketString(packet_data, &index);
  handleFileDownload(pid, client_id, proc_id, taskid, file_path);
  break;

case SERVER_REQ_FILE_UPLOAD:
  std::string src = getPacketString(packet_data, &index);
  std::string dst = getPacketString(packet_data, &index); 
  handleFileUpload(pid, client_id, proc_id, taskid, src, dst);
  break;
\end{verbatim}
\caption{Example of hallucination. While analyzing a Linux/Sshdinjector malware \cite{sshdinjector-ai}, the AI invented 2 botnet commands, SERVER\_REQ\_FILE\_DOWNLOAD and SERVER\_REQ\_FILE\_UPLOAD. In reality, these commands do not exist, there is only a \textit{file transfer} command which supports transfer both ways. The hallucination was spotted because the answer to different question were not coherent. So, the analyst checked the assembly to be certain.}
\label{fig:hallucination}
\end{figure}

\begin{figure}
  \scriptsize
  \begin{verbatim}
printf("%s MAC %02x:%02x:%02x:%02x:%02x:%02x\n", 
        ifa->ifa_name, 
        mac[0], mac[1], mac[2], 
        mac[3], mac[4], mac[5]);
\end{verbatim}
  \caption{Exaggeration example. While analyzing a Linux/Sshdinjector \cite{sshdinjector-ai} malicious sample, the AI reported the code as ``Contains MAC address manipulation code''. This is far fetched, the code simply prints a MAC address, it does not \textit{manipulate} it.}
  \label{fig:manipulation}
\end{figure}

\begin{figure}
  \scriptsize
 \begin{lstlisting}
int main(int argc, char **argv) {
  ...
  int i;
  for (i = 0; i < argc - 1; i++) {
    newArgv[i] = strdup(argv[i + 1]);
    if (strlen(newArgv[i]) == 0)
     memset(newArgv[i], 0, strlen(newArgv[i]));
  }
    
  newArgv[i] = "kdevtmpfs";
  if (i == 0) exitWithError();
  return 0;
}    
\end{lstlisting}
  \caption{This is the presumed source code for Linux/Devura \cite{devura}, generated by ChatGPT 4. The AI forgot to include in code lots of information: creation of a socket, use for forkpty, handling program argument etc. Omissions are frequent and happen for all AI}
  \label{fig:omission}
\end{figure}

\subsection{Speed of analysis}
\label{sec:speed}

\subsubsection{Speed of Local AI}
\label{sec:local-ai}

Remote generative AI usually respond to question in a few seconds.
For local AI, we tested queries on 3 different hosts:

\begin{enumerate}
\item A \textbf{laptop} with Intel Core i7 of 11th generation, 32G RAM, no usable GPU.
\item An \textbf{intermediate} cloud instance with 2x 4-core Intel Xeon (Skylake), 32 GB RAM, no GPU.
\item A \textbf{high} performance cloud instance with 4x 6-core AMD EPYC 7413, 120G RAM, with GPU NVIDIA GA102GL A40.
\end{enumerate}

For local AI, only queries to small binaries or simplified crackme programs finish in a reasonable time on the laptop. Queries to decompile malware do not respond in a reasonable lapse of time. See Table \ref{tab:local-time}.
Using the intermediate cloud instance is still insufficient, we need to raise the bar with a high performance instance to get fast answers.

\begin{table*}[h]
\begin{tabular}{p{2cm}|p{5cm}|p{3cm}|p{4cm}}
    \textbf{Host} & \textbf{Model} & \textbf{Question} & \textbf{Result} \\
    \hline
    Laptop & QuantFactory/granite-8b-code-instruct-4k-GGUF & Write Hello World & $<$ 10 sec \\
    % Write Golang Hello World in max 4 lines
    Laptop & QuantFactory/granite-8b-code-instruct-4k-GGUF & Decompile Goldoon function with main functionalities (fcn.00001028) & $>$ 5 min \\
    % 400 instructions
    Laptop & DeepSeek r1 params=1.5b & Goldoon main & Failure: no decompiled code \\
    Laptop & hhao/qwen2.5-coder-tools params=7b & Goldoon main & $>$ 5 min \\
    Intermediate & deepseek-r1:32b & Goldoon main & 27 minutes. Did not decompile. Only text explanations. Not usable. \\
    Intermediate & codellama:34b & Goldoon main & 9 minutes. Wrong. Not usable. \\
    Intermediate & codestal:latest & Goldoon main & 6 minutes. Not usable. \\
    Intermediate & dolphin3:latest & Goldoon main & 2 minutes. Not usable \\
    Intermediate & phi4:latest & Goldoon main & 7 minutes. Not usable \\
    Intermediate & hhao/qwen2.5-coder-tools:14b & Goldoon main & $>$ 30 minutes. Error: empty response. \\
    \textbf{High} & hhao/qwen2.5-coder-tools:32b & Goldoon main & \textbf{43 seconds. Usable.} \\
    \end{tabular}
\caption{Time for local AI to answer to given questions, based on host specifications. The table shows running local models require a high performance server.}
\label{tab:local-time}
\end{table*}

\subsubsection{Time gain/time loss for analysis}
\label{sec:time-gain}

In terms of speed, we tried to measure time gain or time loss induced by using AI.

Measuring time to analyze binaries with precision is impossible: (1) we don't know how much effort researchers spent \cite{devura} \cite{goldoon} \cite{rudedevil-analysis}, (2) our own AI-assisted analysis was slowed down by the very fact of writing conference articles, and (3) time to analyze a binary obviously depends on researcher's skills, if s/he has already seen a similar sample, and the desired level of details.

Therefore, we estimate an approximate time to complete the analysis, based on our years of experience at reversing binaries. On one side, we compare the time we spent with AI to mark all points of Tables \ref{tab:compare-devura}, \ref{tab:compare-goldoon} and \ref{tab:compare-rudedevil}. For Goldoon, we failed to easily discover the remote URL (see table \ref{tab:compare-goldoon}), consequently we add to our analysis time, the time we spent to debug the issue.
On the other side, we measure estimated time for a reverse engineer to mark all ticks (those found by the initial human only analysis, but also those only reported by AI). All timings should be understood as approximate. They take in account the time it takes to \textit{think} (reverse engineers often let their minds drift when they hit a hard point and come back to it later, thinking about it in background). A variation of 1 day is entirely possible. See Table \ref{tab:time-gain}.

\begin{table*}[h]
  \begin{tabular}{c|c|c|c}
    Malware & Estimated Human Only Time & AI-Assisted Time & Conservative Time loss/save \\
    \hline
    Devura & 3-4 days & 2 days  & \textbf{1-2 days} \\
    Goldoon & 4-5 days & 2-4 days  & \textbf{0.5 day} \\
    RudeDevil & 5-6 days & 3-4 days & \textbf{1-3 days} \\
  \end{tabular}
  \caption{Estimated approximate time for a reverse engineer to analyze a given malware, with or without AI, for the same quality of analysis. All results are in favour of AI assistance, but the time save varies from one sample to another.}
  \label{tab:time-gain}
\end{table*}

\section{Conclusion}

R2ai enhances the r2 disassembler with AI assistance. The tool has several options which make the queries to the AI very customizable and suitable for AI-assisted reverse engineering. 

For reverse engineering, so far, best results have been obtained with a paid API key giving access to Anthropic Claude 3.5 and 3.7 Sonnet. Using local AI is possible in theory, but requires a powerful host with several cores, RAM and GPU. With Claude, r2ai can be used efficiently to decompile functions and explain them. The quality of the generated code is excellent (clear, wise naming, good comments etc) but reverse engineers need to keep in mind the code might not cover all cases (AI often omits what it thinks are details).

Hallucinations or exaggerations are also relatively frequent. They are the worse dangers of AI-assisted analysis, can mislead an unsuspecting engineer and require constant attention and critical mind to be detected. Despite its drawbacks, our measures indicate that AI-assisted analysis allows engineers to go faster \textit{and} in more details in their analysis.

\section{Acknowledgments}

We wish to thank Sergi Alvarez, radare2 author for his implementation, reactivity and help. We also express our gratitude to Pr Ludovic Apvrille, for his detailed review of the paper, and several discussions on AI.


\begin{thebibliography}{9}
\bibitem{aidapal}
  Aidapal GitHub repository \url{https://github.com/atredispartners/aidapal}

\bibitem{ida-assistant}
  IDA-Assistant GitHub repository.
  \url{https://github.com/stuxnet147/IDA-Assistant}
  
\bibitem{reveng}
  Reveng.AI \url{https://reveng.ai/}

\bibitem{ghidrassist}
  GhidraAssist GitHub repository \url{https://github.com/jtang613/GhidrAssist}

\bibitem{gpthidra}
  E. Evyatar, 
  \textit{Introducing GptHidra: The AI-Powered Code Assistant for Ghidra}.
  January 2023.
  https://evyatar9.medium.com/introducing-gpthidra-the-ai-powered-code\--assistant-for-ghidra-78844d2bc227
  

\bibitem{binarychat}
  BinaryChat \url{https://github.com/Protosec-Research/BinaryChat}

\bibitem{sidekick}
  SideKick \url{https://sidekick.binary.ninja/}

  \bibitem{ghidra-mcp}
  L. Kirk (Laurie Wired) \textit{Now AI Can Reverse Malware}, March 26, 2015 \url{https://www.youtube.com/watch?v=u2vQapLAW88}


\bibitem{rudedevil-obfuscation}
  A. Apvrille, \textit{Insomni'hack 2025: de-obfuscating strings in Linux/RudeDevil}, March 2025, \url{https://asciinema.org/a/708621}

\bibitem{insomnihack}
  A. Apvrille, \textit{Malware analysis with R2AI}, Insomni'hack, March 2025, \url{https://github.com/cryptax/talks/blob/master/Insomnihack-2025/r2ai.pdf}

\bibitem{devura}
  S. Friedberg, \textit{Unveiling ``sedexp'': A Stealthy Linux Malware Exploiting udev rules}, August 2024 \url{https://www.aon.com/en/insights/cyber-labs/unveiling-sedexp}

\bibitem{goldoon}
  C. Lin and V. Li, \textit{New ``Goldoon'' Botnet Targeting D-Link Devices}, May 2024 \url{https://www.fortinet.com/blog/threat-research/new-goldoon-botnet-targeting-d-link-devices}

\bibitem{rudedevil-analysis}
  R. Groenewoud, \textit{Betting on Bots: Investigating Linux malware, crypto mining, and gambling API abuse}, September 2024 \url{https://www.elastic.co/security-labs/betting-on-bots}

\bibitem{sshdinjector-ai}
  A. Apvrille, \textit{Analyzing ELF/Sshdinjector.A!tr with a human and artificial analyst}, February 2025, \url{https://www.fortinet.com/blog/threat-research/analyzing-elf-sshdinjector-with-a-human-and-artificial-analyst}

\bibitem{prometei-ai}
  A. Apvrille, \textit{Reversing a Prometei botnet binary with r2 and AI}, February 2025, 3 parts, \url{https://cryptax.medium.com/reversing-a-prometei-botnet-binary-with-r2-and-ai-part-one-3cdb3dc6ffab}

\bibitem{r2mcp}
  Radare2 MCP server, \url{https://github.com/radareorg/radare2-mcp}

  
\end{thebibliography}
\end{document}